\documentclass[12pt]{article}
\usepackage{graphicx}
\usepackage{amsmath,amsfonts}

\def\,{\ifmmode\mskip\thinmuskip\else\leavevmode\thinspace\fi}

\newcommand{\be}{\begin{equation}}
\newcommand{\ee}{\end{equation}}
\newcommand{\ba}{\begin{eqnarray}}
\newcommand{\ea}{\end{eqnarray}}
\newcommand{\nn}{\nonumber }
\newcommand{\bm}[1]{\langle #1\rangle}
\newcommand\unit[1]{\,\textmd{#1}}
\def\bc{\begin{center}} \def\ec{\end{center}}
\newcommand{\dd}{{\mbox{d}}}

\newcommand{\vecc}[1]{\mathbf{#1}}
\newcommand{\vecs}[1]{\mathbf{#1}{\lower-.2em\hbox{}^{2}}}
\newcommand{\veck}[1]{\mathbf{#1}{\kern-.2em\hbox{}^{2}}}

\def\lsi{\raise0.3ex\hbox{$<$\kern-0.75em\raise-1.1ex\hbox{$\sim$}}}
\def\gsi{\raise0.3ex\hbox{$>$\kern-0.75em\raise-1.1ex\hbox{$\sim$}}}
\newcommand{\lsim}{\mathop{\lsi}}

\title{The lepton pair production in heavy ion collisions
in perturbation theory}
\author {E.~Barto\v s$^{1,2}$,
S.~R.~Gevorkyan$^{1,3}$, E.~A.~Kuraev$^1$ and N.~N.~Nikolaev$^4$}
\date{} 

\begin{document}
\maketitle
\begin{center} {
$^{1}$ \it Joint Institute of Nuclear Research, 141980 Dubna, Russia\\
$^{2}$ \it Dept. of Theor. Physics, Comenius Univ., Bratislava,
Slovak Republic \\
$^{3}$ \it  Yerevan Physics Institute, 375036 Yerevan, Armenia\\
$^{4}$ \it Institut f. Kernphysik, Forshchungszentrum J\"ulich,
D-52425 J\"ulich, Germany and L.~D.~Landau Institute for
Theoretical Physics, 142432 Chernogolovka, Russia\\
}
\end{center}

\vspace*{2cm}

\begin{abstract}
We derive the first terms in the amplitude of lepton pair
production in the Coulomb fields of two relativistic heavy ions.
Using the Sudakov technique, which very simplify the calculations
in momentum space for the processes at high energies, we  get the
compact analytical expressions for differential cross section of
the process under consideration in the lowest order in fine
structure constant (Born approximation) valid for any momentum
transfer and in a wide kinematics region  for produced particles.
Exploiting the same technique we consider the next terms of
perturbation series (up to fourth order in fine structure
constant) and investigate their energy dependence and limiting
cases. It has been shown that taking in account all relevant terms
in corresponding order one obtains the expressions which are gauge
invariant and finite. We estimate the contribution of the Coulomb
corrections to the total cross section and discuss the
cancellations of the different terms which holds in the total
cross section.
\end{abstract}
\vspace*{2cm}

\section{Introduction}

\begin{figure}[ht]
\begin{center}
\includegraphics[scale=1.]{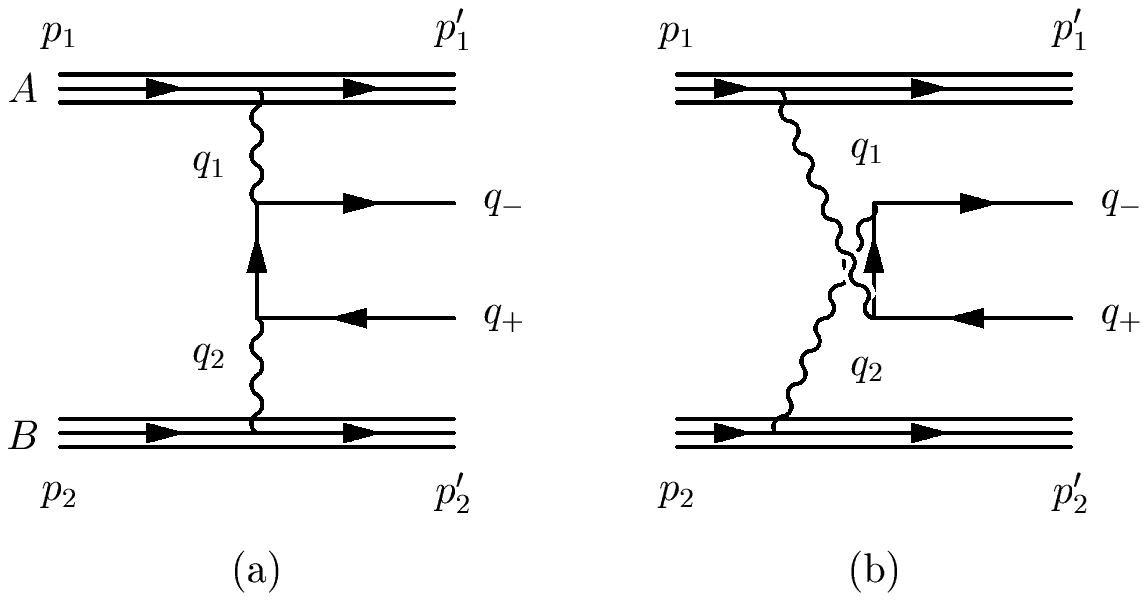}
\caption{The lowest order Feynman diagrams for the two--photon
pair production in the process $A+B\rightarrow A+B+e^+e^-$.}
\label{fig:1}
\end{center}
\end{figure}

The process of lepton pair production in collision of relativistic
heavy ions attracted a rising interest in the past years which is
connected mainly with  operation of  ~RHIC (~Lorentz factor
$\gamma=E/M=100$) and LHC ($\gamma= 3000$). The total cross
section of the production of lepton pair in the collision of two
relativistic particles
\ba \label{eq:1} A+B\rightarrow e^+e^-+ A+B \ea
in its lowest order in fine structure constant
$\alpha=\frac{e^2}{{4\pi}}= \frac{1}{137} $ has been known long
ago \cite{LL,R}. Over the last decade a lot of works have been
done to investigate this process for the case when $A$ and $B$ are
heavy ions (see e.g. \cite{LMS} and references there in). When
heavy ions collide at relativistic velocities their
electromagnetic fields are ~Lorentz contracted and the process
(\ref{eq:1}) can be viewed as pair production in two external
fields which can be represented by classical electromagnetic
potentials. Such approach allows \cite{HTB1,HTB2} to investigate
the impact parameter dependence of this process  and calculate its
cross section with higher accuracy, than predicted by widely
exploited Weizs\"acker-Williams (WW) approximation. This works are
based on the Born approximation (see Fig.~\ref{fig:1}) but even in
this case the obtained expressions are complex, so that the
certain "state of art" efforts in their computing are required
\cite{GWUSE}.

\begin{figure}[hb]
\begin{center}
\includegraphics[scale=1.]{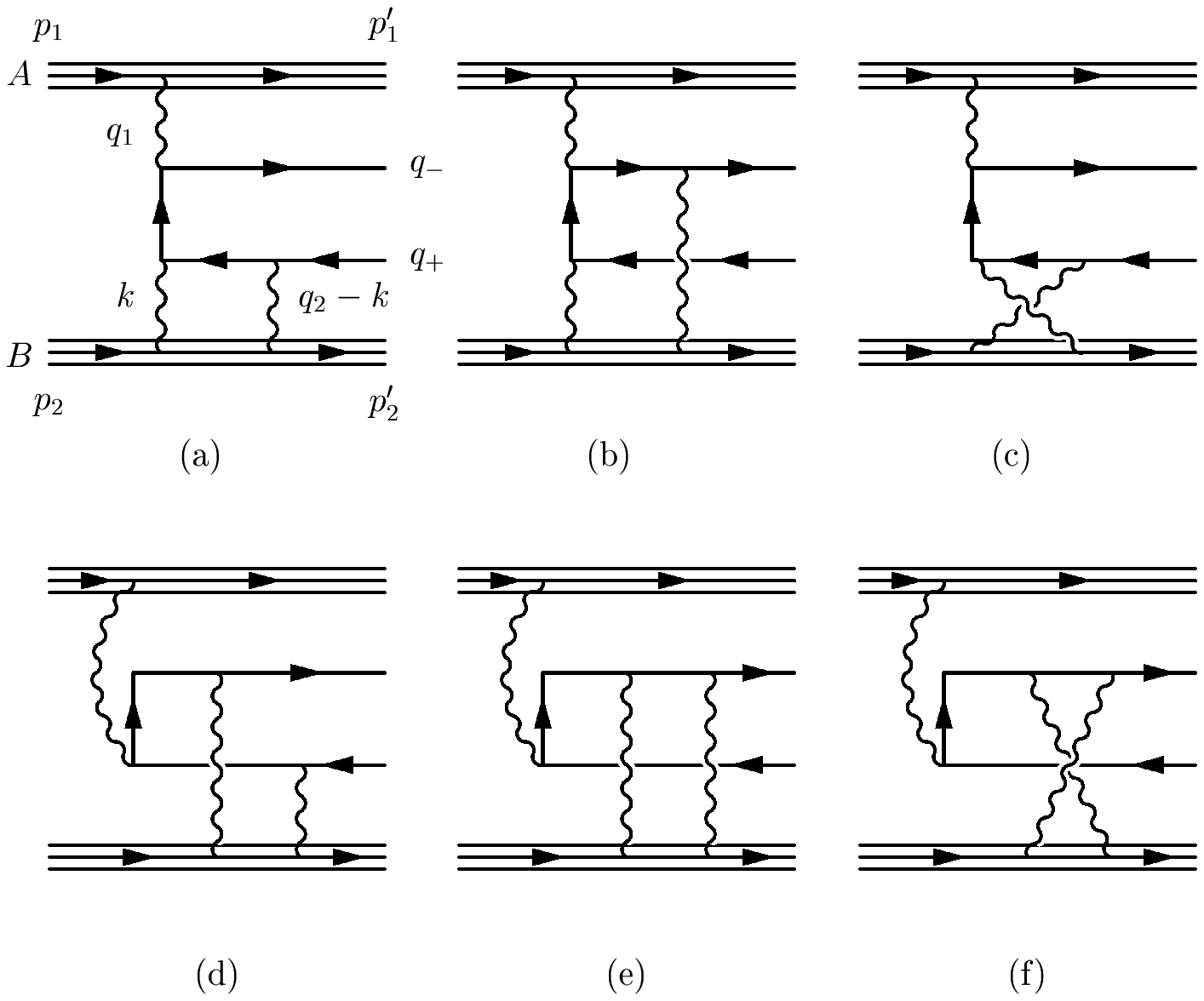}
\caption{Gauge invariant set of Feynman diagrams for spin--triplet
production which determines the matrix element $M_{(2)}^{(1)}$.}
\label{fig:2}
\end{center}
\end{figure}

The topical problem is to take into account the Coulomb
corrections (CC) to the amplitude of this process (see
Fig.~\ref{fig:2},~\ref{fig:3}). In spite of the fact that CC are
determined by the powers of fine structure constant $\alpha$, it
enters the amplitude in combination $(\alpha Z_1)^{n_1}(\alpha
Z_2)^{n_2}(\alpha Z_1Z_2)^{n_3} $, where $Z_1,Z_2$ are the
colliding ions charge numbers, $n_1+n_2\geq 2$ is the number of
exchanged photons between nuclei and the produced pair, $n_3\geq
0$ is the number of photons exchanged between the colliding
nuclei. For heavy ion collisions the parameter $\alpha Z$ is of
the order of unity and the problem arise  to take into account the
effect of CC. The exact knowledge of total yield of lepton pairs
in heavy ion collisions is important issue, because the pair
production consist a huge background in experiments with
relativistic heavy ions. For example the total cross section for
the process (1) at ~RHIC energies is tens kilobarns for the heavy
ion collisions. At LHC energies this quantity becomes a hundreds
kilobarns according to the Racah formula \cite{R}. Moreover, the
intensive pair production  can destroy the ion beams circulating
in the accelerator as a result of electron capture by heavy ion
(see e.g. \cite{BB}).

Recently the problem of CC in the process (1) was investigated in
the number of papers \cite{SW,BM,BGMP,ERG} with the unexpected
result. The authors \cite{BM,BGMP} solved the Dirac equation  for
an electron in the Coulomb field of two highly relativistic
nuclei. Having used the crossing symmetry property, they connected
the amplitude for electron scattering in the Coulomb fields of two
heavy ions with amplitude for pair production\footnote{The similar
trick was exploited many years ago \cite{M} to get the probability
of pair production in the condensed media from the probability of
the electron bremsstrahlung in the matter.}. As a result the total
cross section of the process (1) turn out to coincide with its
Born approximation, so that CC haven't impact on the total lepton
pairs yield. The same result have been obtained in the work
\cite{ERG} in which the authors use the eikonal approximation for
electron scattering at high energies. They took advantage of the
common wisdom  that the interaction of lepton pair with Coulomb
fields of two highly relativistic nucleus can be represented as a
product of the eikonal amplitudes for interaction of electron and
positron with a nucleus $A$ and $B$ separately. Such approach
allows one to take into account the contribution of CC in
amplitude of the process (\ref{eq:1}) with the quoted above
result: the total cross section exactly coincide with the Born
term. This works induced the series of critical papers
\cite{ISS,LM}, where it was explained why this result is so
unexpected and how assumptions which have been done through its
derivation  can affect this result. As was observed in \cite{ISS}
the main contribution to the CC in the process (\ref{eq:1}) comes
from the case when one of the colliding ions radiate a single
photon and the produced pair interact with another nucleus by
means of the arbitrary number of photons. In fact the
contributions of such type can be obtained with the use of common
WW approximation, which leads to appearance of large logarithms in
energy for such terms, which can't be cancelled by the next order
of CC where such logarithms are absent. On the other side it has
been shown \cite{LM} that the change in the order of integration
in regularized integrals (the Coulomb phase for electron
scattering diverges, so it needs regularization) completely change
the final result. In  \cite{BGMP} the two approaches employed in
investigation of this problem were analyzed in detail, with the
conclusion that the perturbation approach in fact correspond to
the exclusive process of the coherent production of lepton pairs
in the collisions of relativistic ions (i.e. reaction
(\ref{eq:1})). As to the approach based on retarded solutions of
the Dirac equation it gives the inclusive cross section for
production of arbitrary number of the lepton pairs, i.e. in this
way one obtain the contribution of inclusive pairs production. The
reason of such unexpected result from our point of view is the
inapplicability of crossing symmetry property to the higher order
(beyond the Born approximation) terms, the fact, which is known
long ago (see e.g. \cite{KLS}) Thus the issue of calculation of
the CC in the correct way in the process (\ref{eq:1}) is an open
task up to now and it turn out to be more complex problem that it
seems from the first glance. All this stimulated us to investigate
the CC contribution in the perturbative expansion of amplitude for
exclusive process under discussion in more details, with the aim
to understand how one can take into account the CC in heavy ion
collisions in the framework of perturbative QED.

Our paper is organized as follows. In the next section 2 we
present the amplitude for the lepton pair creation in the
collision of relativistic particles in the Born approximation. We
are restricted by kinematics of pionization region which gives the
main contribution to the cross section, so that the pair
production doesn't influence the kinematics of the colliding
particles. Moreover, we supposed that momenta transfer between the
produced pair and initial particles are much smaller in comparison
with the energy of initial particles, e.g. we consider the case of
their quasi--elastic scattering. We use the powerful Sudakov
technique for the evaluation of the Feynman diagrams in momentum
representation which allows to obtain the results with power
accuracy in energy. The compact analytical expressions have been
obtained  for the Born amplitude and differential cross section of
the process (1) which allows one to calculate it in the wide
interval of kinematic variables.In the limiting case of quasi-real
photons the well known results of WW approximation can be easily
obtained from this expressions.

In the section 3 we consider all possibilities for pair production
by three photon exchange, so called amplitudes of the type
$M_{(2)}^{(1)}$. We present the general formula for its
contribution to the singlet triplet pair production amplitude and
relevant cross section. Our result appears to be the
generalization of the known result for the amplitude of the
process of pair photoproduction in the odd charge state derived in
\cite{GPS}. We have done the numerical estimate for the relevant
contribution to the total cross section in the main (double
logarithmic) approximation.

In the section 4 we present  the results of the amplitude
computation and corresponding contributions in the cross section
from processes of two photon exchange between each ion and the
created pair (fourth order terms in fine structure constant) and
the case, when one of the ions is attached to the pair by single
photon and another ion is connected with a pair by three photons
in all possible ways. In the last case, our result appears to be
generalization of the known result in the case of photoproduction
of the pair in even charge state \cite{IM}. We show that in any
order the obtained expressions are gauge invariant and finite. We
analyze the main contributions from obtained CC to the total cross
section and show that the number of remarkable cancellation
between the different CC and their interference with the Born term
take place.

In the section 5 we shortly listed the main results obtained in
the paper.

In the Appendix~A we give the results of analytical integration on
loop momenta for spin--singlet and spin--triplet parts of CC,
which allows one to obtain the first CC in the analytical form.

In the Appendix~B we present the spin structures which determines
the fourth order terms in the amplitude of the process under
consideration.

\section{The Born approximation}
We are interested in the exclusive coherent process of lepton pair
production in the collision of two relativistic nuclei with charge
numbers $Z_1,Z_2$ \ba \label{eq:2} A(p_1)+B(p_2) \to
e^+(q_+)+e^-(q_-)+A(p_1')+B(p_2').\ea We define the usual c.m.s.
total energy  of colliding nuclei $s=(p_1+p_2)^2 = 4\gamma_1
\gamma_2 m_1m_2$ where $m_i$  and  $\gamma_i=\frac{E_i}{m_i}$ are
the masses and Lorentz factors  of the colliding nuclei. Keeping
in mind the fast decrease of $\gamma^*\gamma^* \to e^+e^-$ cross
section with the momentum transfer, i.e. the photons virtuality
and with the invariant mass  of the pair $s_p=(q_++q_-)^2 $ we
will work in the kinematics
$$s=(p_1+p_2)^2\gg| q_1^2|,\;|q_2^2|,\;s_p.$$
Later on we calculate  the amplitudes  neglecting the pieces of
the order  $\frac{m_i^2}{s}$, so our results are valid for
relativistic particles with power accuracy. The total cross
section of the reaction (\ref{eq:2}) in the lowest order in fine
structure constant (Born approximation) is known long time ago
\cite{LL,R}. Nevertheless we carry out the derivation of amplitude
and differential cross section in this approximation using the
Sudakov technique, which is very useful for calculation of Feynman
diagrams (FD) for the processes at high energies.This allow us  to
illustrate  the method and approximations which we will use later
on in the  CC derivation. Moreover we will obtain the compact
analytical expressions, which are valid not only for the heavy ion
collisions, but for the lepton pair production in the Coulomb
fields of any relativistic particles (hadrons, leptons).

The lowest order FD describing the process (\ref{eq:2}) are
depicted in the Fig.~\ref{fig:1}. The corresponding amplitude read
\begin{gather} \label{eq:3}
M_{(1)}^{(1)}=-i\frac{(4\pi\alpha)^2 Z_1 Z_2}{{q_1^2q_2^2}}\bar
u(p_1') \gamma_\mu u(p_1) \bar u(p_2')\gamma_\nu
u(p_2)g_{\mu\alpha}g_{\nu\beta}\bar u(q_-) T_{\alpha\beta}
v(q_+)\\ \nn T_{\alpha\beta}=\gamma_\beta \frac{{\hat q_1-\hat
q_++m}}{{(q_1-q_+)^2-m^2}}\gamma_\alpha+\gamma_\alpha \frac{{\hat
q_2-\hat q_++m}}{{(q_2-q_+)^2-m^2}}\gamma_\beta.
\end{gather}
Using the Gribov's decomposition of metric tensor $
g_{\mu\nu}=g_{\mu\nu}^{\bot}+\frac{2}{s}( p_1^\mu p_2^\nu+p_1^\nu
p_2^\mu) $ it is easy to show that at high energies the main
contribution  is given by its longitudinal part $
g_{\mu\alpha}g_{\nu\beta}\to \left(\frac{2}{s}\right)^2 p_1^\nu
p_2^\mu p_1^{\alpha} p_2^{\beta}.$ As a result the expression
(\ref{eq:3}) can be cast in the form
\begin{gather}
\label{eq:4}
M_{(1)}^{(1)}=-is\frac{{(8\pi\alpha)^2Z_1Z_2}}{{q_1^2q_2^2}}\bar
u(q_-)R_{(1)}^{(1)} v(q_+)N_1N_2, \\ \nonumber R_{(1)}^{(1)}=
\frac{1}{s}p_1^\mu p_2^\nu T_{\mu\nu},\quad N_1=\frac{1}{ s}\bar
u(p_1')\hat{p_2}u(p_1),\quad N_2=\frac{1}{s}\bar
u(p_2')\hat{p_1}u(p_2).
\end{gather}
For spinless projectile and target as well as for the colliding
fermions at definite chirality state $|N_1|=| N_2|=1$, so that
later on\footnote{The spin of the colliding nuclei is unimportant
because of the s--channel helicity conservation in the interaction
of relativistic particles with Coulomb field.} we will set
$N_1=N_2=1$. It should be noted that if we consider the heavy
ion's dimensions it is enough to multiply the amplitude
(\ref{eq:4}) by the colliding ion's formfactors $F(q_i^2)$.

In what follows we will use the standard Sudakov expansion of all
the momenta in the two almost light-light vectors $\tilde p_1$, $
\tilde p_2 $ and the two--dimensional transverse component $q_\bot
p_1=q_\bot p_2=0 $
\begin{gather} \label{eq:5}
q_1=\alpha_1\tilde p_2+\beta_1\tilde p_1+q_{1\bot},\quad
q_2=\alpha_2\tilde p_2+\beta_2\tilde p_1+q_{2\bot},\quad
q_{\pm}=\alpha_{\pm}\tilde p_2+\beta_{\pm}\tilde p_1+q_{\pm\bot}, \\
\nn  \tilde p_1=p_1-p_2\frac{p_1^2}{s},\quad \tilde
p_2=p_2-p_1\frac{p_2^2}{s},\quad \tilde p_1^2=\tilde
p_2^2=O\left(\frac{m^6}{s^2}\right),\quad s \approx 2p_1p_2\approx
2\tilde p_1\tilde p_2.
\end{gather}
From the on--mass shell condition, for instance
${p_1'}^2=(p_1-q_1)^2= m_1^2$, ${p_2'}^2=(p_2-q_2)^2=m_2^2$, one
can easily see that $  \beta_1\sim \alpha_2\sim\frac{m}{{\sqrt
s}}$, $ \alpha_1 \sim\beta_2\sim\frac{q^2}{s}$ which allows to
neglect, where it is possible, the values of $\alpha_1$, $\beta_2$
which greatly simplify calculations and allows to obtain the
results with the power accuracy. In this approximation we have
\begin{gather} \label{eq:6}
q_1\approx\beta_1\tilde p_1+ q_{1\bot},\quad
q_2\approx\alpha_2\tilde p_2+q_{2\bot},\\ \nn -q_1^2=\frac{\veck{
q_1} +\beta_1^2 m_1^2}{1-\beta_1} ,\quad -q_2^2=\frac{\veck{
q_2}+\alpha_2^2m_2^2}{1-\alpha_2}\\ \nn
s_p=(q_++q_-)^2=(q_1+q_2)^2\approx \alpha_2\beta_1 s-(\vecc{q_1}+
\vecc{q_2})^2.
\end{gather}
Hereafter $\vecc{q_i}$ always stand for $q_{i\bot}$ etc. The
conservation laws provides the relations among the introduced
variables and thus the limits on $\beta_1$, $\alpha_2$ variation
\begin{gather} \label{eq:7}
\beta_1=\beta_-+\beta_+,\quad
\alpha_2=\alpha_-+\alpha_+,\quad \vecc{q_1}+\vecc{q_2}=\vecc{
q_-}+\vecc{ q_+}, \\ \nn \frac{s_p+(\vecc{ q_1}+\vecc{
q_2})^2}{s}<\beta_1\sim\alpha_2<1.
\end{gather}
As we shall see later the production amplitudes takes a simple
form in terms of the  following variables
\ba \label{eq:8}
z_{\pm} = \beta_{\pm}/\beta_1,\qquad z_-=1-z_+,\qquad \vecc{
k_{\pm}}=\vecc{ q_{\pm}}-z_{\pm}\vecc{ q_1}.
\ea
In what follows we will work in the kinematics of pionization
region (similar approach was used in \cite{KNZ}), which give the
main contribution to the total cross cross section of the process
under consideration. For such kinematics the following limitations
take place $$\frac{m^2}{s}\ll\alpha_2\approx\beta_1\ll 1.$$ As a
result the expression (\ref{eq:4}) can be represented as
\begin{equation}  \label{eq:9}
M^{(1)}_{(1)}=-is\frac{(8\pi\alpha)^2
N_1N_2}{(\veck{q_1}+\beta_1^2m^2_1) (\veck{q_2}+\alpha_2^2m^2_2)}
Z_1 Z_2 \bar{u}(q_-)R^{(1)}_{(1)}v(q_+).
\end{equation}
It is convenient to use  the gauge invariance condition
${q_1}^\mu T_{\mu\nu}=0$, which in view of  (\ref{eq:4}) entails

\begin{equation} \label{eq:10}
p_1^\nu T_{\nu\mu}=-\frac{q_{1\bot}^{\nu}}{\beta_1}
T_{\nu\mu}\equiv -\frac{|\vecc{ q_1}|}{\beta_1}e_1^\nu
 T_{\nu\mu}, \quad e^\nu_1=\frac{\vecc{ q_1}}{|\vecc{ q_1}|}.
\end{equation}
Substituting this relation in (\ref{eq:4}) and after a bit of
Dirac algebra we find

\begin{gather} \label{eq:11}
R^{(1)}_{(1)}=\frac{|\vecc{q_1}|}{s\beta_1}
\left[m\hat{e}_1R_1+\hat{e}_1\hat{Q}_1+2z_{+}\vecc{Q_1}\vecc{e_1}
+2|\vecc{q_1}|z_{+}z_{-}R_{1}\right]\hat{p}_2 \\ \nn
R_1=S(\vecc{k_-})-S(\vecc{k_+}),\quad \vecc{ Q_1} =\vecc{k_-}
S(\vecc{ k_-})+\vecc{ k_+}S(\vecc{ k_+}),\quad S(\vecc{
k})=\frac{1}{ \vecc{k}^2 + z_+z_-\veck{ q_1} + m^2}.
\end{gather}

This representation for the amplitude is manifestly gauge
invariant, i.e. its vanishes when $q_1\rightarrow 0 $. Doing in
(\ref{eq:11}) the replacement
$\vecc{q_1}=\vecc{q_+}+\vecc{q_-}-\vecc{q_2}$ it is easy to see
that it vanishes also when $\vecc{q_2}\to 0$.

For completeness we cite here the alternative expression for the
amplitude which can be obtained if one use the  gauge invariance
constraint by another photon  ${q_2}^\mu T_{\mu\nu}=0 $.
\begin{gather} \label{eq:12}
R^{(1)}_{(1)}=\frac{|\vecc{q_2}|}{s\alpha_2}\left[m\hat{e}_2R_1+
\hat{e}_2\hat Q_1+2 y_+\vecc{
Q_1}\vecc{e_2}+2|\vecc{q_2}|y_+y_-R_1 \right]\hat{p}_1,\\\nn
R_1=S(\vecc{l_-})-S(\vecc{l_+}),\quad\vecc{Q_1}=\vecc{l_+}S(\vecc{l_+})
+\vecc{l_-} S(\vecc{l_-}),\\\nn
S(\vecc{l})=\frac{1}{m^2+\vecc{q_2}^2y_+y_-+\vecc{l}^2},\quad
\vecc{l_\pm}=\vecc{q_\pm}-y_\pm\vecc{q_2},\quad
y_\pm=\frac{\alpha_\pm} {\alpha_2},\quad y_++y_-=1.
\end{gather}
The two factors
$$\frac{|\vecc{q_1}|}{\veck{q_1}+m_1^2
\beta_1^2},\quad\frac{|\vecc{q_2}|}{\veck{q_2}+m_2^2
\alpha_2^2}$$ are the familiar bremsstrahlung amplitudes for the
photons with transverse momenta $\vecc{q_{1,2}} $, which
determines the luminosity of $\gamma\gamma$ collisions \cite{BB}.
The square of $\bar u(q_-) R_{(1)}^{(1)} v(q_+) $ (up to the well
known kinematics factors) gives the cross section of the process
$\gamma^*\gamma^*\to e^+e^-$.

After a lengthy but straightforward calculation we obtain
\begin{multline} \label{eq:13}
\frac{1}{4}\sum{\Big|\bar u(q_-)R_{(1)}^{(1)}v(q_+)\Big|}^2=
\frac{1}{4}
Sp (\hat q_-+m)R^{(1)\ast}_{(1)}(\hat q_+-m) R_{(1)}^{(1)}=\\
\frac{1}{2} z_+z_- \veck{q_1} \left[(m^2 + 4z_+z_-\veck{ q_1})
R_1^2 + (z_+^2+z_-^2)\veck{Q_1} + 4z_+z_-(z_+-z_-)R_1 \vecc{
q_1}\vecc{ Q_1}\right].
\end{multline}
The different terms in this expression have a transparent physical
meaning in terms of the transverse and scalar (often called
longitudinal) polarizations of the virtual photon $\gamma^{*}_{1}$
and there is one-to-one correspondence with the discussion of
helicity structure function in diffractive DIS \cite{NPZ} and
diffractive production of vector mesons \cite{KNZ}. Namely in the
expression (\ref{eq:13}) the first term  proportional to the $m^2$
corresponds to pair production by transverse photon with the sum
of the lepton helicities $\lambda+\bar{\lambda}=\pm 1$, the  term
$(z_+^2+z_-^2)\veck{Q_1} $ provided the excitation of the pair
with $\lambda+\bar{\lambda}=0 $ by the transverse photons. The
term $\propto 4z_+z_-\veck{q_1} R_1^2 $ in (\ref{eq:13}) describes
the pair production by scalar photons, whereas the last term in
(\ref{eq:13}) is the ST (LT) interference which vanishes upon the
phase space integrations by $z_+$ and $z_-$.

Because of the exact conservation of the helicity of relativistic
electrons in a Cou\-lomb scattering, the above structure of the
amplitude (\ref{eq:11}) and its square (\ref{eq:13}) will be
retained also in higher order. This important property  will be
widely used later on constructing the terms of higher orders in
fine structure constant.

The differential cross section of the process (\ref{eq:2}) is
connected with the full amplitude by the relation
\ba \label{eq:14}
\dd\sigma=\frac{1}{{8s}}\sum{| M|}^2\dd\Gamma.
\ea
The summation is carried over the final particles polarization.
The phase space volume for four particles in the final state can
be expressed through the variables introduced above and phase
volume of the final leptons
\ba \label{eq:15}
\dd\Gamma_{\pm}&=&\frac{\dd^3q_+}{{2E_+}}\frac{\dd^3q_-}{ {
2E_-}}\delta^4(q_1+q_2-q_+-q_-)\ea in the following way
\begin{align} \label{eq:16} \dd\Gamma
&=\frac{1}{(2\pi)^8}\frac{\dd^3p_1'}{2E_1'}\frac{\dd^3p_2'}{2E_2'}
\frac{\dd^3q_-}{2E_-}\frac{\dd^3q_+}{2E_+}
\delta^4(p_1+p_2-p_1'-p_2'-q_+-q_-)\\\nn &=\frac{{\dd
s_p}}{{4(2\pi)^8
s(1-\alpha_2)(1-\beta_1)}}\frac{\dd\beta_1}{\beta_1}
\dd^2\vecc{q_1}\dd^2\vecc{q_2}\dd\Gamma_{\pm}.
\end{align}
In obtaining this expression we used the following relations for
the final particles momenta
$$\frac{{\dd^3q}}{ E}=2\delta(q^2-m^2)\dd^4q=s \dd\alpha \dd\beta
\dd^2\vecc{q}\delta(s\alpha \beta-\vecc{q}^2-m^2).$$ The above
expressions allows one to calculate the differential cross section
of the process under consideration in the Born approximation with
power accuracy in energy.

The total cross section of pair production in the elementary
process $\gamma^{*}_1\gamma^{*}_2 \to e^+e^-$ drops rapidly as
soon as one of the photons is off--mass shell, which imposes the
upper limit of integration over the photons transfer momenta
$\veck{q_1},\veck{q_2} \lsim m^2$. This limits are much smaller
than the characteristic momentum where the formfactors are at work
$$
|\vecc{q_{1,2}}|^2 \lsim m_{e}^2 \ll \frac{6 }{ \langle R_{ch}^2
\rangle_{A}} \approx \frac{10 }{ R_A^2} \approx \frac{0.25}{
A^{2/3}}\unit{GeV}^2,
$$
where $R_A \sim 1.2 A^\frac{1}{3}\unit{fm}$ is the nuclear radius,
so that the effect of nuclear formfactors can be safely neglected.
(It is not the case for the production of muon pairs or hadronic
states though.)

For this reason the evaluation of the total cross section in the
case of electron pair production can be done in the point like
nucleus approximation, i.e. using the cited above expressions. It
can be  shown that integrating the expression for the differential
cross section (\ref{eq:14}) using the relations (\ref{eq:9}),
(\ref{eq:13}) and (\ref{eq:16}) one  obtains the famous Racah
formula \cite{R} for the total cross section
\ba \label{eq:17}
\sigma =\frac{\alpha^4Z_1^2Z_2^2}{\pi m^2}\frac{28}{27} \Big[L^3
-2.2 L^2 +3.84 L -1.636\Big]
\ea
where $ L=\ln(\gamma_1\gamma_2) $.
The total cross section of the process under consideration in the
Born approximation is the growing function of the invariant energy
(as the third power of logarithm in energy). This is the direct
consequence of the fact that both photons are quasi real, which
bring to second power of logarithm in (\ref{eq:17}) and one power
come from the common for all orders integration by rapidity
variable $\beta_1$.
Before going to calculation of properly Coulomb corrections let us
do the following remark. The lepton pair created by two photons is
produced in charge even i.e. spin--singlet state. The exchange of
the odd number of photons between the Coulomb fields of heavy ions
and produced leptons will lead to a pair creation in charge odd
i.e. spin--triplet state, which doesn't interfere with Born
amplitude. Nevertheless later on we have to take into account only
the interference of the Born amplitude with the amplitude where
the four photon exchanges take place between leptons and colliding
particles (see Fig.~\ref{fig:3}). Now we can go to the main topic
of present work, calculations of Coulomb corrections (CC). We
begin from the simplest case which are provided by three photons
exchange between the produced lepton pair and colliding particles.

\section{The  Coulomb corrections. The spin-triplet\\
(C-odd state) pair production}
We start with the discussion of the triplet pair production,which
proceeds via the exchange of the odd number of photons between
colliding particles and produced  pair. The amplitude relevant to
first CC of such type consist of two parts
$M_{(12)}=M_{(2)}^{(1)}+M_{(1)}^{(2)}$, where the amplitude
$M_{(2)}^{(1)}$ corresponds to the case when the lepton pair is
attached to colliding particle $A$ by one photon and connects with
$B$ through two photons in all possible ways. The case when two
photons are attached to $A$ (one to $B$) is described by the term
$M_{(1)}^{(2)} $. As would be shown later this term can be
obtained one from another by simple substitution, so it is enough
to consider one of them.

The amplitude  $M_{(2)}^{(1)}$ is described by six Feynman
diagrams  which are depicted in the Fig.~\ref{fig:2}. The only
essential difference from the Born approximation is the additional
photon exchange with four--momentum $ k=\alpha\tilde
p_2+\beta\tilde p_1+ k_\bot$ by which the integration has been
done in the amplitude. To integrate over Sudakov variables
$\alpha$ and $\beta$ we use twelve FD instead of six which is
relevant to this amplitude and introduce the statistical factor
$\frac{1}{ {2!}}$. This trick permits us to provide the eikonal
type integration over $\beta$ using the identity
\ba \label{eq:18}
\int{\frac{\dd\beta}{ {2\pi
i}}\left(\frac{1}{{s\beta+i0}}+\frac{1}{{-s\beta+i0}}\right)}=1.
\ea
The convergence of the integration over $\alpha$ is provided by
all six FDs obtained by permutation of all absorption points of
exchanged photons by the pair. Closing the $\alpha$ integration
contour to the upper half--plane one can see that only four of
them are relevant.

Using the same approximation as in the Born case we obtain
\begin{align} \label{eq:19}
M_{(2)}^{(1)}&=s\frac{(4\pi\alpha)^3 Z_1
Z_2^2}{2\pi(\veck{q_1}+m_1^2
\beta_1^2)}\bar u(q_-)\Re_{(2)}^{(1)} v(q_+), \\\nn
\Re_{(2)}^{(1)}&=\frac{|\vecc{q_1}|}{s\beta_1}\int\frac{\dd^2
\vecc{k}}{\pi}\frac{1}{\vecc{k}^2(\vecc{q_2}-\vecc{k})^2}
\left[m\hat e_1R_2(k)+\hat{e}_1\hat{Q}_2(k)\right.\\\nn &\quad
\left. +2z_{+}\vecc{Q_2(k)}\vecc{ e_1} +2|\vecc{q_1}|z_{+}z_{-}
R_{2}(k)\right]\hat p_2\\
&\equiv \frac{|\vecc{q_1}|}{ s\beta_1}
\left[m\hat{e}_1R_{(2)}^{(1)}+\hat{e}_1\hat{Q}_{(2)}^{(1)}+2z_{+}
\vecc{Q_{(2)}^{(1)}}\vecc{e_1} +2|\vecc{
q_1}|z_{+}z_{-}R_{(2)}^{(1)}\right]\hat p_2, \nn
\end{align}
where in close analogy to amplitudes of diffractive DIS \cite{KNZ}
\begin{gather}  \label{eq:20}
R_2(\vecc{ k})=S(\vecc{ k_-})+S(\vecc{ k_+}) - S(\vecc{k_-}-\vecc{
k})- S(\vecc{ k_+}-\vecc{ k}), \\\nn \vecc{ Q_2}(\vecc{ k})=\vecc{
k_-}S(\vecc{ k_-})-\vecc{ k_+}S(\vecc{ k_+})-(\vecc{ k_-}-\vecc{
k}) S(\vecc{ k_-}-\vecc{ k})+(\vecc{ k_+}-\vecc{ k})S(\vecc{
k_+}-\vecc{ k}).
\end{gather}
One can see that the spin structure of first CC is precisely the
same as in the Born case. This is the direct result of lepton
helicity conservation and as was shown in \cite{IM,IKSS} is valid
to all orders of perturbation series for the case, when one
consider the single photon exchange between the pair and
projectile $A$ with any number of exchanges with target $B$ and
vice versa. Using the relations from Appendix A one can do the
integration in (\ref{eq:19}) and obtains the analytical expression
for $\Re_{(2)}^{(1)}$. In general case this expression is rather
cumbersome therefore we will present in the Appendix A only the
result of integration in the case, when the single photon which is
attached to projectile $A$ is real $(\vecc {q_1}\to 0)$. The
square of amplitude $M_{(2)}^{(1)}$  can be obtained  by analogy
with the Born case
\begin{multline}  \label{eq:21}
\frac{1}{4}\sum{\Big|\bar u(q_+)
\Re_{(2)}^{(1)}v(q_-)\Big|}^2=\frac{1}{ 2}z_+z_-
\veck{q_1}\bigg[\left(m^2 + 4z_+z_-\veck{
q_1}\right)(R_{(2)}^{(1)})^2  \\
\left.+(z_+^2+z_-^2)\left(\Big(\vecc{ Q_{(2)}^{(1)}}\Big)^2 +
4z_+z_-(z_+-z_-)R_{(2)}^{(1)}\vecc{ q_1}\vecc{
Q_{(2)}^{(1)}}\right)\right].
\end{multline}
The amplitude $M^{(2)}_{(1)}$ ( two--photon exchange between the
lepton pair and the nucleus $A_1$) is readily obtained from
$M^{(1)}_{(2)}$ by following substitutions
\ba  \label{eq:22}
\beta_1 \leftrightarrow \alpha_2,\quad z_{\pm} \leftrightarrow
 y_{\pm},\quad \vecc{q_1} \leftrightarrow \vecc{q_2},\quad \vecc{e_1}
\leftrightarrow \vecc{e_2},\quad \hat p_2 \leftrightarrow \hat
p_1.
\ea
Let us estimate the considered CC contribution to the total cross
section. This can be easily done if one restricted  to the  terms
which grows with energy as the main power of the logarithm. In
this case the  CC behavior is defined by the second power of large
logarithm  $L=\ln(\gamma_1\gamma_2)$ i.e.
$$ L^2=\int\limits_{\beta_{1 min}}^{1}{\frac{\dd\beta_1}{\beta_1}}
\int\frac{\dd^2 \vecc{q_1}}{\pi}
\frac{\veck{q_1}}{(\veck{q_1}+m_1^2\beta_1^2)^2},\quad
\beta_{1 min}=\frac{1}{\gamma_1\gamma_2},$$
so that to the leading logarithmical accuracy
\begin{gather} \label{eq:23}
\sigma_{odd}^{tot}=\frac{\alpha^6Z_1^2Z_2^2(Z_1^2+Z_2^2)L^2}{\pi
m^2}P,\\\nn P=\int\frac{\dd^2\vecc{q_-}}{\pi}\int\frac{\dd^2
\vecc{q_+}}{\pi}\left[ m^4\Big(R_{(2)}^{(1)}\Big)^2+
\frac{2}{3}m^2\Big(\vecc{Q_{(2)}^{(1)}}\Big)^2\right]_{\vecc{
q_1}=0}.
\end{gather}
Using the expressions for $R_{(2)}^{(1)}$ and
$\vecc{Q_{(2)}^{(1)}}$ from Appendix A we calculated this quantity
with the result $ P\approx 3,6 $. We want to emphasize that the
general expression (\ref{eq:19}) contains the leading logarithms
contribution as well as the nonleading ones omitting only terms
which are suppressed by factor $\frac{m^2}{ s}$ and vanish in the
high energy limit.

The considered case $\vecc{q_1}\to 0$ corresponds to WW
approximation i.e. to the case when the scattered ion $A$ is not
detected and will scatter at very small angles. We see that for
the case with semi-inclusive setup (when the scattered ion is
fixed) we have much more complicated picture, which nevertheless
can be described by cited above expressions.

The contribution  from the interference of the amplitudes
$M^{(1)}_{(2)}$ and $M^{(2)}_{(1)}$ is enhanced only by the first
power of large logarithm  L due to the boost effect and its impact
on the total cross section will be consider further.Here we only
cite this interference contribution to the charge odd part of the
total cross section in the general form
\ba  \label{eq:24}
\sigma^{int}_{odd}=\frac{\alpha^6 Z_1^3Z_2^3 L}{\pi^2}
\int\frac{\dd^2\vecc{q_1}}{\pi
\veck{q_1}}\frac{\dd^2\vecc{q_2}}{\pi \veck{q_2}}\int \dd s_p\int
\dd\Gamma_\pm
\frac{1}{4}Sp(\hat{q}_-+m)\Re^{(2)}_{(1)}(\hat{q}_+-m)
\tilde{\Re}^{(1)}_{(2)}.
\ea
\section{The Coulomb corrections to the spin singlet \\(C-even)
pair production}

\begin{figure}[hb]
\begin{center}
\includegraphics[scale=1.]{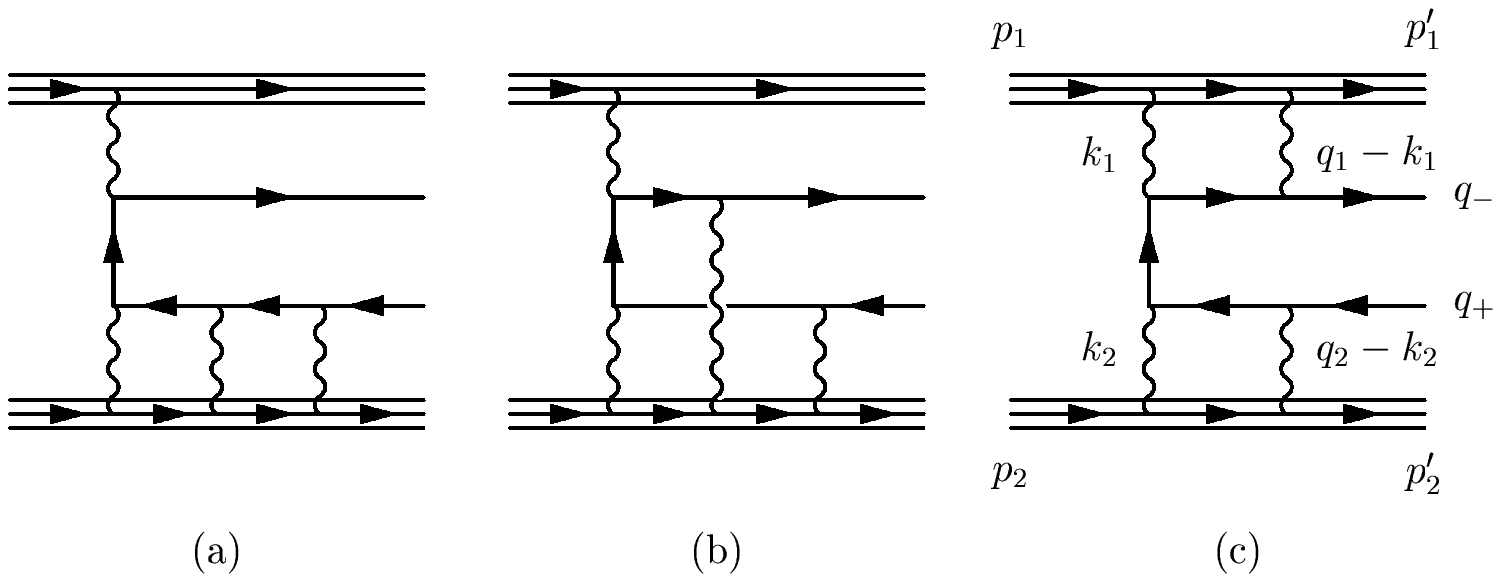}
\caption{Some Feynman diagrams for the four--photon pair
production in the process $A+B\rightarrow A+B+e^+e^-$.}
\label{fig:3}
\end{center}
\end{figure}

The spin singlet pair production CC are provided by four photons
exchange between the colliding ions and produced pair in all
possible ways. Conventionally one can divided them in two set. The
first one is determined by FD some of which are depicted on
Fig.~\ref{fig:3}~a, b. The corresponding amplitude denotes by
$M_{13}=M_{(3)}^{(1)}+M_{(1)}^{(3)}$ represent the sum of all
possible single attachments to one of the colliding ions with
three photons exchanges between the lepton pair and another ion.
The second set is determined by FD depicted on Fig.~\ref{fig:3} c
and include all double photon exchanges between lepton pair and
every colliding ions. This amplitude we write as $M_{(2)}^{(2)}$.
Using the same approximations and technique as above for the
amplitude $M_{(3)}^{(1)}$ we obtain
\begin{align}
\label{eq:25}
M_{(3)}^{(1)}&=\frac{is}{3!}\frac{(4\pi\alpha)^4Z_1Z_2^3}
{(2\pi)^2 (\veck{ q_1}+\beta_1^2m_1^2)}\bar
u(q_-)\Re_{(3)}^{(1)}v(q_+), \\ \nn
\Re_{(3)}^{(1)}&=\frac{|\vecc{q_1}|}{s\beta_1}\int\frac{\dd^2
\vecc{k_1}}{\pi} \frac{\dd^2\vecc{k_2}}{\pi}\frac{1}{\vecs{
k_1}\vecs{ k_2} (\vecc{ q_2}-\vecc{ k_1}-\vecc {k_2})^2}\\ \nn
&\quad\times\left[m\hat e_1R_3(k)+\hat e_1\hat Q_3(k)+2z_+ \vecc{
Q_3}(k) \vecc{ e_1} +2|\vecc{ q_1}|z_+z_-R_3(k)\right]\hat p_2\\
\nn &\equiv \frac{|\vecc {q_1}|}{s\beta_1} \left[m\hat
e_1R_{(3)}^{(1)}+\hat e_1\hat Q_{(3)}^{(1)}+2z_{+} \vecc{
Q_{(3)}^{(1)}}\vecc{ e_1}| +2|\vecc{
q_1}|z_+z_-R_{(3)}^{(1)}\right]\hat p_2,
\end{align}
where in close analogy to amplitudes of diffractive DIS
\begin{align} \label{eq:26}
R_3(\vecc{ k_1},\vecc{ k_2})&=-S(\vecc {k_-})+S(\vecc{
k_-}-\vecc{k_1})+S(\vecc{ k_-}-\vecc{ k_2})
+S(\vecc{k_-}-\vecc{q_2}+\vecc{k_1}+\vecc{ k_2})\\
&\quad -\left[\vecc{k_-}\leftrightarrow \vecc{k_+}\right],\nn \\
\nn \vecc{ Q_3}(\vecc{ k_1},\vecc{ k_2})&=\vecc{ k_-}S(\vecc{
k_-})-(\vecc{k_-}-\vecc{k_1})S(\vecc{k_-}-\vecc{k_1})-(\vecc{k_-}-
\vecc{k_2})S(\vecc{k_-}- \vecc{k_2})\\
& \quad -(\vecc{k_-}-\vecc{q_2}+\vecc{k_1}+\vecc{k_2})
S(\vecc{k_-}-\vecc{q_2}+\vecc{k_1}+\vecc{k_2}) -
\left[\vecc{k_-}\leftrightarrow \vecc{k_+}\right]. \nn
\end{align}
One can see that as in the case of third order CC considered above
the fourth order amplitude  of first kind (i.e. with one single
attachment) has the same spin structure as the Born one and can be
obtained from it by simple replacement $(R_1,\vecc{Q_1})
\leftrightarrow (R_3,\vecc{ Q_3})$. To get the amplitude
$M_{(1)}^{(3)}$ it is enough to do the replacement (\ref{eq:22})
in the above expressions. It is easy to check that the amplitude
$M_{12}$ vanish when any of photons momenta $q_1$ or $q_2$ goes to
zero, so that the gauge invariance is provided. Finally, in the WW
limit $ q_1,q_2\to 0 $ one obtains from (\ref{eq:25}),
(\ref{eq:26}) the expression which coincide with respective one
from \cite{GI}.

The main contribution to the total cross section come from the
interference of the $M_{13}$ with the Born amplitude
$M_{(1)}^{(1)}$. To the leading logarithmic accuracy  this
interference can be read
\begin{gather}  \label{eq:27}
\sigma^{(3)}_{(1)}+\sigma^{(1)}_{(3)}=-\frac{2\alpha^6(Z_1Z_2)^2
(Z_1^2+Z_2^2)L^2}{\pi m^2}P_1, \\ \nn
P_1=\int\frac{\dd^2\vecc{q_2}}{\pi}\frac{\dd^2\vecc{q_-}}{\pi}
\frac{m^2}{\veck{q_2}} \Bigg[m^2R_1R_{(3)}^{(1)}+\frac{2}{3}\vecc{
Q_1}\vecc{ Q_{(3)}^{(1)}}\Bigg]_{\vecc{ q_1}=0}.
\end{gather}
Compare this expression with the cross section in the case of
C-odd CC (\ref{eq:23}) one can see that the both terms are of the
same order in fine structure constant and have the same energy
dependence. Moreover we make a point that $P_1=P $. Indeed, the
product $R_1R_{(3)}^{(1)}$ contains 16 terms and making the proper
shifts of the integration variables, one can establish one-to-one
correspondence to the 16 terms in $\Big(R_{(2)}^{(1)}\Big)^2$.
Similarly the same can be done with the vector products
$\vecc{Q_1}\vecc{Q_{(3)}^{(1)}}$ and $\Big(\vecc{
Q_{(2)}^{(1)}}\Big)^2$. This can be regarded as a manifestation of
the well known AGK rules \cite{AGK}. Really, the two photon
exchange production of lepton pair can be viewed as a diffraction
excitation of the photon, whereas the 3--photon exchange can be
viewed as an absorption correction to the one--photon exchange.
Furthermore, the total cross section for pair photoproduction in
the Coulomb field reads \cite{DBM}
\ba \label{eq:28}
\sigma^{\gamma Z\to
e^+e^-Z}(s)=\frac{28Z^2\alpha^3}{9m^2}\left[\ln\frac{s}
{m^2}-\frac{109}{42}-(Z\alpha)^2\sum_{n=1}^{\infty}\frac{1}{n(n^2+
(Z\alpha)^2)}\right],
\ea
which for the relevant order gives
$$ P=P_1=\frac{28}{9}\zeta(3) \approx 3.6,\quad
\zeta(3)=\sum_{n=1}^{\infty}\frac{1}{n^3}=1.202. $$ As was
mentioned above this result was verified by numerical
calculations.

Returning to the second set of the fourth order CC (two photons
exchange between the pair and every heavy ion) we calculated all
respective terms in accordance with the FD (see Fig.~\ref{fig:3}
c) and obtained the matrix element $M^{(2)}_{(2)}$ in the
following form
\begin{gather} \label{eq:29}
 M^{(2)}_{(2)}=is\alpha^42^4(Z_1Z_2\pi)^2\bar u(q_-)
 \Re^{(2)}_{(2)}v(q_+),\\ \nn
\Re^{(2)}_{(2)}=\int\frac{\dd^2\vecc{k_1}}{\pi}
\int\frac{\dd^2\vecc{k_2}}{\pi}
\frac{1}{\vecs{k_1}(\vecc{q_1}-\vecc{k_1})^2}
\frac{1}{\vecs{k_2}(\vecc{q_2}-\vecc{k_2})^2}
(1+P_{ud})R^{(2)}_{(2)}.
\end{gather}
The structure $\Re^{(2)}_{(2)}$ has so called up--down symmetry,
which is decoded in the permutation operator $P_{ud}$, which acts
as follows
\ba \label{eq:30}
P_{ud}F(p_1,p_2,\alpha,\beta,q_1,k_1,q_2,k_2)=
F(p_2,p_1,\beta,\alpha,q_2,k_2,q_1,k_1).
\ea
The quantity $R^{(2)}_{(2)}$ is the sum of all  terms
corresponding to all possible FD for this case
\ba \label{eq:31}
R^{(2)}_{(2)}=R_{1243}+R_{1324}+R_{2431}+
R_{3124}+R_{1342}+R_{1423}.
\ea
The structures  $R_{igkl}$ are cited in Appendix B. Unfortunately
we do not succeed in trying to write down the result for spin
structures in this case as it has be done above for the terms
where there is a single attachment of exchanged photon to one of
the colliding ions.

Nevertheless we cite here the leading contribution to the C--even
state pair creation cross section which come from the interference
of this amplitude and the Born term
\ba \label{eq:32}
\sigma^{int}_{even}=\frac{\alpha^6(Z_1Z_2)^3
L}{\pi^2}\int\frac{\dd^2\vecc{q_1}}{\pi
\veck{q_1}}\frac{\dd^2\vecc{q_2}}{\pi \veck{q_2}}\int \dd s_p \int
\dd \Gamma_\pm
\frac{1}{4}Sp(q_-+m)\Re^{(2)}_{(2)}(q_+-m)\tilde{\Re}^{(1)}_{(1)}.
\ea
This expression is the same order in fine structure constant as
the considered above interference between the Born amplitude and
first set of fourth order term (see (\ref{eq:27})), but it has a
one power less in its energy dependence.

\begin{figure}[hb]
\begin{center}
\includegraphics[scale=1.]{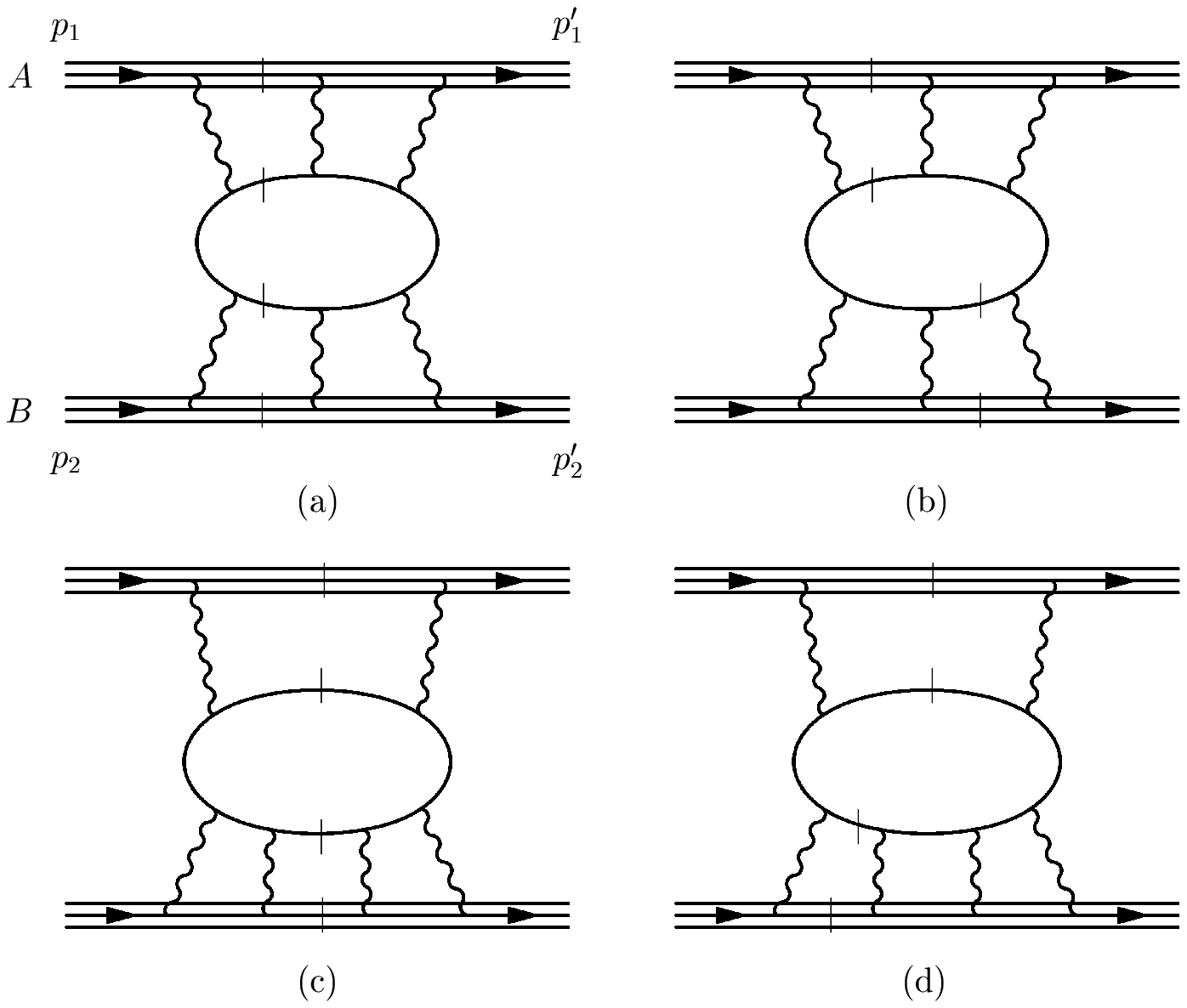}
\caption{Different s--chanel cuts for forward elastic scattering
in sixth order of perturbation theory.} \label{fig:4}
\end{center}
\end{figure}

On the other side the first power dependence on L and the same
order in $\alpha$ has contribution to the total cross section
arising from the interference between the amplitudes $
M_{(2)}^{(1)}$ and $ M_{(1)}^{(2)}$ (see (\ref{eq:24})) which
corresponds to pair production in C-odd state. From very general
arguments  this contributions have to cancel each other within the
logarithmical accuracy. This statement\footnote{We are grateful to
L.~Lipatov for discussion on this issue.} follows from the fact
that the relevant (sixth order in $\alpha$) elastic amplitude for
$AB$ scattering at zero angle (see Fig.~\ref{fig:4}) has only
single logarithmical enhancement factor due to the boost freedom
of a light--light scattering block, whereas the s--channel
discontinuity of this amplitude has no logarithmical enhancement
at all. Therefore in the total cross section of pair production
the terms proportional to $ (Z_1Z_2 \alpha^2)^3 L $ are absent. It
is interesting to check this statement by straightforward
calculations, which will be done elsewhere. Moreover it seems very
natural that the contributions of the form
$(Z_1\alpha)^{2k+1}(Z_2\alpha)^{2l+1}$ are absent in the total
cross section in any orders of perturbation theory. This statement
is in the complete agreement with the results recently obtained
for the Coulomb corrections \cite{LM2}.

\section{Concluding remarks}

In this work we have considered lepton pair production in the
Coulomb field of two highly relativistic heavy ions. Using the
powerful technique of Sudakov variables,which allows one to obtain
the results which are correct with the power accuracy in energy,
we get the first terms in perturbation expansion of amplitude (up
to the fourth order in fine structure constant) and show that they
are finite and gauge invariant. We obtain the simple analytical
expressions for the Born amplitude and relevant cross section,
which allows one to calculate the lepton pair yield in the wide
interval of kinematic variables. It is shown that the terms in the
amplitude, which corresponds to FD with at least one single
exchange between created pair and one of the ion, have the same
spin structure in all orders in $\alpha$ and we proposed the
simple rules to build them.

In every order of perturbation theory we analyze  the obtained CC
isolating  the leading in energy terms and show that remarkable
cancellations among the different terms in the total cross section
take place. Finally the numerical estimates of the main
contribution from the CC in the total cross section have been
done.

\section*{Acknowledgements}
We are grateful to Stanislav Dubni\v{c}ka for the participation in
the initial stage of this work and to Valery Serbo for useful
discussions. The work is supported by INTAS 97-30494. E.~B.
and E.~K. have support from SR-2000 grant.
E.~K. is grateful to the participants of DESY Theory seminar for
interesting discussions.

\newpage
\appendix
\section*{Appendix A}
\renewcommand{\theequation}{A.\arabic{equation}}
\setcounter{equation}{0}

We give here the result of two--dimensional integration by loop
momenta for scalar and vector structures determining the first CC
\begin{align}
I(\vecc{q_-})&=
\int\frac{\dd^2\vecc{k}}{\pi}\frac{1}{(\vecc{k}^2+\lambda^2)
((\vecc{q}-\vecc{k})^2+\lambda^2)((\vecc{q_-}-\vecc{k})^2+m^2)}\\
&=  \frac{1}{\vecc{q}^2}\left[\frac{1}{\bm{q_-}}+
\frac{1}{\bm{q_+}}\right]
\ln\frac{\vecc{q}^2}{\lambda^2}+\frac{1}{\Delta}
\left[\frac{\veck{q_-}}{\bm{q_-}}+\frac{\veck{q_+}}
{\bm{q_+}}-1\right]\ln\frac{m^2}{\vecc{q}^2}  \nn \\ & \quad +
A_1\ln\frac{\bm{q_-}}{m^2}+A_2\ln\frac{\bm{q_+}}{m^2},  \nn
\end{align}
\begin{align}
&A_1=-\frac{\vecc{q}\vecc{q_-}}{\vecc{q}^2}\left[\frac{1}{\Delta}+
\frac{1}{\bm{q_-}\bm{q_+}}\right]+\frac{1}{\Delta}
\left[\frac{\veck{q_-}}{\bm{q_-}}+\frac{\veck{q_+}}{\bm{q_+}}-
\frac{m^2\vecc{q}\vecc{q_+}}{\bm{q_-}\bm{q_+}}\right],\\ \nn
&A_2=-\frac{\vecc{q}\vecc{q_+}}{\vecc{q}^2}
\left[\frac{1}{\Delta}+\frac{1}{\bm{q_-}\bm{q_+}}\right] +
\frac{1}{\Delta}\left[\frac{\veck{q_-}}{\bm{q_-}}+\frac{\veck{q_+}}
{\bm{q_+}}-\frac{m^2\vecc{q}\vecc{q_-}}{\bm{q_-}\bm{q_+}}\right],
\nn
\end{align}
\begin{align}
\vecc{I}(\vecc{q_-})&=\int\frac{\dd^2\vecc{k}}{\pi}\frac{\vecc{k}}
{(\vecc{k}^2+\lambda^2)((\vecc{q}-\vecc{k})^2+\lambda^2)
((\vecc{q_-}-\vecc{k})^2+m^2)}\\
&=\frac{\vecc{q}}{\vecc{q}^2\bm{q_+}}\ln\frac{\bm{q_+}}{\lambda^2}-
\frac{1}{\Delta} \vecc{q_+}\left[\ln\frac{\bm{q_-}}{\vecc{q}^2}+
\ln\frac{\bm{q_+}}{m^2}\right]  \nn \\
&\quad +\frac{1}{\Delta}\vecc{q}\left[\frac{\veck{q_+}}{\bm{q_+}}
\ln\frac{\bm{q_+}}{m^2}-\frac{\bm{q_-}-\vecc{q}^2}{\vecc{q}^2}
\ln\frac{\bm{q_-}}{\vecc{q}^2}\right], \nn
\end{align}
where we use the notation $\langle a\rangle=\vecc{a}^2+m^2$ and
besides
$$
\Delta=\bm{q_-}\bm{q_+}-m^2\vecc{q}^2,\quad\vecc{q}=\vecc{q}_++\vecc{q}_-.
$$
Using these integrals and the expression (\ref{eq:20}) we obtain
for scalar combination which enters $M^{(1)}_{(2)} $\begin{align}
R_{(2)}^{(1)}&=\int\frac{\dd^2\vecc{k}}{\pi}\frac{R_{2}(k)}
{\vecc{k}^2(\vecc{q}-\vecc{k})^2} \\ \nn
&=\frac{2}{\Delta}\left(1-\frac{\veck{q_-}}{\bm{q_-}}-
\frac{\veck{q_+}}{\bm{q_+}}\right)\ln\frac{\bm{q_-}\bm{q_+}}
{m^2\vecc{q}^2}+\frac{2}{\vecc{q}^2}\left(\frac{1}{\bm{q_+}}-
\frac{1}{\bm{q_-}}\right)\ln\frac{\bm{q_-}}{\bm{q_+}}, \\
R_{2}(k)&=\frac{1}{\bm{q_-}}-\frac{1}{\bm{k-q_-}}+
\frac{1}{\bm{q_+}}-\frac{1}{\bm{k-q_+}}. \nn
\end{align}
For the case of the vector structures we obtain
\begin{align}
\vecc{Q_{(2)}^{(1)}}&=\int\frac{\dd^2\vecc{k}}{\pi}\frac{
\vecc{Q_{2}}(k)}{\vecc{k}^2(\vecc{q}-\vecc{k})^2}\\ \nn
&=(\vecc{q_-}-\vecc{q_+})\left[\frac{m^2}{\Delta}\left(\frac{1}
{\bm{q_-}}+\frac{1}{\bm{q_+}}\right)\ln\frac{\bm{q_-}\bm{q_+}}
{m^2\vecc{q}^2}+\frac{1}{\vecc{q}^2}\left(\frac{1}{\bm{q_-}}-
\frac{1}{\bm{q_+}}\right)\ln\frac{\bm{q_-}}{\bm{q_+}}\right]\\ \nn
&\quad+\frac{\vecc{q}}{\vecc{q}^2}\left[(\bm{q_+}-\bm{q_-})
\left(\frac{1}{\Delta}\ln\frac{\bm{q_+}\bm{q_-}}{m^2\vecc{q}^2}
-\frac{1}{\bm{q_+}\bm{q_-}}\ln\frac{m^2}{\vecc{q}^2}\right)\right.\\
\nn &\quad+\left.
2\left(\frac{1}{\bm{q_+}}\ln\frac{\bm{q_+}}{m^2}-
\frac{1}{\bm{q_-}}\ln\frac{\bm{q_-}}{m^2}\right)\right], \\\nn
\vecc{Q_{2}}(k)&=\frac{\vecc{q_-}}{\bm{q_-}}+\frac{\vecc{k}-\vecc{q_-}}
{\bm{k-q_-}}-\frac{\vecc{q_+}}{\bm{q_+}}-
\frac{\vecc{k}-\vecc{q_+}}{\bm{k-q_+}}.
\end{align}

\section*{Appendix B}
\renewcommand{\theequation}{B.\arabic{equation}}
\setcounter{equation}{0}
We cite here the spin structures which determines the fourth order
 amplitude of second class (see expression (\ref{eq:32}))
\begin{align}
&R_{1243}=\frac{1}{s}\cdot\frac{\hat{p}_1(\hat{q}_--\hat{q}_1+m)
\hat{p}_2}{\frac{\beta_+}{\beta_-}\bm{q_-}+\bm{q_--q_1}},
\\ \nn &R_{1324}=-\frac{1}{s^2}\cdot
\frac{\hat{p}_1(\hat{q}_--\hat{k}_1+m)\hat{p}_2(\hat{q}_--\hat{k}_1
-\hat{k}_2+m)\hat{p}_1(-\hat{q}_++\hat{q}_2-\hat{k}_2+m)\hat{p}_2}
{\frac{\beta_+}{\beta_-}\bm{q_-}\bm{q_--k_1-k_2}+\bm{q_--k_1}
\bm{-q_++q_2-k_2}},\\ \nn
&R_{2431}=\frac{1}{s^2}\cdot\frac{\hat{p}_1(\hat{q}_--\hat{q}_1+
\hat{k}_1+m)\hat{p}_2(-\hat{q}_++\hat{k}_1+m)\hat{p}_1}
{\alpha_-\bm{-q_++k_1}+\alpha_+\bm{q_--q_1+k_1}},\\ \nn
&R_{3124}=-\frac{1}{s^2}\cdot\frac{\hat{p}_2(\hat{q}_--\hat{k}_2+m)
\hat{p}_1(-\hat{q}_+\hat{q}_2-\hat{k}_2+m)\hat{p}_2}
{\beta_+\bm{-q_-+k_2}+\beta_-\bm{q_--q_1-k_2}},\\ \nn
&R_{1423}=-\frac{1}{s^2}\cdot
\frac{\hat{p}_1(\hat{q}_--\hat{k}_1+m)\hat{p}_2(-\hat{q}_++\hat{q}_1
+\hat{k}_2-\hat{k}_1+m)\hat{p}_1(-\hat{q}_++\hat{k}_2+m)\hat{p}_2}
{\frac{\beta_+}{\beta_-}\bm{q_-}\bm{-q_++q_1+k_1-k_2}+
\bm{q_--k_1}\bm{-q_++k_2}},\\ \nn &R_{2413}=-\frac{1}{s^2}\cdot
\frac{\hat{p}_1(\hat{q}_--\hat{q}_1+\hat{k}_1+m)\hat{p}_2(-\hat{q}_+
+\hat{k}_1+\hat{k}_2+m)\hat{p}_1(-\hat{q}_++\hat{k}_2+m)\hat{p}_2}
{\frac{\beta_+}{\beta_-}\bm{q_-}\bm{-q_++k_2+k_1}+\bm{-q_++k_2}
\bm{q_2-q_++k_1}}.\nn
\end{align}
The symmetry property is caused by the charge-conjugation symmetry
of the amplitude. There are 24 FD contributing to $M^{(2)}_{(2)}$.
Instead of them it is convenient to consider $24\ast 2\ast 2=96$
FD which take as well the permutations of emission and absorption
points of exchanged photons to the nuclei. The result must be
divided by $(2!)^2$. This trick provides the convergence of
$\alpha_1,\beta_2$ integrals. All the 24 FD, describing the
interaction of 4 virtual photons with the pair components are
relevant providing the convergence of $\beta_1,\;\alpha_2$
integrations. When closing the contour of integration say in upper
planes only 12 from them become relevant.

\newpage
\renewcommand{\textfraction}{0}
\end{document}